\documentclass{iaus}
\usepackage{graphicx}
\newcommand{\micron}{$\mu$m}
\title[~~NIR spectroscopy of post-SB galaxies] 
{Direct Constraints on the Impact of TP-AGB Stars on the SED of
  Galaxies from Near-Infrared Spectroscopy}

\author[Zibetti \etal] 
{Stefano Zibetti$^1$, Anna Gallazzi$^1$, St\'ephane Charlot$^2$, Anna
  Pasquali$^3$ \and Daniele Pierini\thanks{Visitor astronomer at the
    Max-Planck-Institut f\"ur extraterrestrische Physik, Garching}}

\affiliation{$^1$Dark Cosmology Centre, Niels Bohr Institute,
  University of Copenhagen, \\ Juliane Maries Vej 30, DK-2100
  Copenhagen, Denmark \\ email: {\tt zibetti, gallazzi @dark-cosmology.dk} \\[\affilskip]
  $^2$Institut d'Astrophysique de Paris, France \\[\affilskip]
  $^3$Astronomisches Rechen-Institut, Heidelberg, Germany
}

\pubyear{2011} 
\volume{284}  
\pagerange{1--12}
\jname{The Spectral Energy Distribution of Galaxies}
\editors{R.J. Tuffs \&  C.C.Popescu, eds.}
\begin{document}

\maketitle

\begin{abstract}
  We present new spectro-photometric NIR observations of 16
  post-starburst galaxies especially designed to test for the presence
  of strong carbon features of thermally pulsing AGB (TP-AGB) stars,
  as predicted by recent models of stellar population
  synthesis. Selection based on clear spectroscopic optical features
  indicating the strong predominance of stellar populations with ages
  between 0.5 and 1.5 Gyr and redshift around 0.2 allows us to probe
  the spectral region that is most affected by the carbon features of
  TP-AGB stars (unaccessible from the ground for $z\sim0$ galaxies) in
  the evolutionary phase when their impact on the IR luminosity is
  maximum.  Nevertheless, none of the observed galaxies display such
  features. Moreover the NIR fluxes relative to optical are consistent
  with those predicted by the original Bruzual \& Charlot (2003)
  models, where the impact of TP-AGB stars is much lower than has been
  recently advocated.

  \keywords{galaxies: general - galaxies: galaxies: stellar content –
    galaxies: fundamental parameters - infrared: galaxies, stars -
    stars: AGB and post-AGB.}
\end{abstract}

\firstsection 
\section{TP-AGB Stars: the big unknown}

The modeling and interpretation of the spectral energy distribution
(SED) of stellar populations via stellar population synthesis (SPS) is
a fundamental tool to understand galaxy properties and their
evolution. Yet we are far from a complete comprehension and a reliable
modeling of some stellar evolutionary phases which strongly affect the
energy output of stellar populations. Among them, the so-called
thermally pulsing - asymptotic giant branch (TP-AGB for short) phase
has been the focus of debate among modelists for several years
(e.g. \cite[Maraston 2005]{maraston_05}; \cite[Bruzual
2007]{bruzual_07}; \cite[Marigo \etal~2008]{marigo+08}). TP-AGB stars
are believed to be major contributors to the NIR flux of stellar
populations with age between 0.5 and 1.5 Gyr. However, quantitative
predictions of the lifetimes, luminosities and emission continuum
shapes and spectral features of these stars and of their influence on
the integrated spectra of even ``simple'' stellar populations, are
still very uncertain and strongly model-dependent. Different
treatments of TP-AGB stars (i.e. prescriptions to model their
lifetimes and luminosities, different stellar spectral libraries) in
popular SPS codes can yield significant differences, in particular
concerning \textit{i)} the flux ratios in different NIR bands and in
NIR vs optical bands and \textit{ii)} the presence (or absence) of
sharp spectral features due to molecules in the atmospheres of those
stars.  The age and spectral ranges affected by TP-AGB stars are of
particular interest for two main applications. \textit{a)} At low
redshift, the NIR is often used as tracer of stellar mass in galaxies,
both unresolved (e.g. \cite[Bell \etal~2003]{bell+03}) and resolved
ones (e.g. \cite[Zibetti, Charlot \& Rix, 2009]{ZCR09}), because of
the (supposed) lower sensitivity to the young stellar components that
often dominate the visible part of the spectrum and because of lower
dust extinction. However, different treatments of TP-AGB stars in SPS
models result in systematic differences in global stellar mass
estimates of several tens of percent and in substantially different
estimates of the stellar mass density contrast of structures like
spiral arms within galaxies (see Zibetti, Charlot \& Rix,
2009). \textit{b)} At $z\gtrsim 2.5$, the bulk of stellar populations
has mean ages in the range 0.5--1.5 Gyr that is (allegedly) mostly
affected by TP-AGB stars. As demonstrated by, e.g., \cite[Maraston et
al. (2006)]{maraston+06} and \cite[Kannappan \& Gawiser
(2007)]{kannappan_gawiser_07}, uncertainties on TP-AGB models severely
affect stellar mass and age determinations for galaxies at those
redshifts (see also \cite[Conroy, Gunn \& White, 2009]{conroy+09}).

\section{Testing SPS model predictions: NIR spectrophotometric
  observations of post-starburst galaxies at $z\sim0.2$}

We directly test the impact of TP-AGB stars on the integrated SED of
galaxies by obtaining spectra in H and K bands, which we combine with
available optical spectroscopy and photometry. Such an SED can be
confronted with SPS predictions and used to constrain model templates
for galaxies at low and high redshifts. In order to maximize the
impact of TP-AGB stars, we target ``post-starburst'' galaxies, whose
star formation an intense burst 0.5-1.5 Gyr back in the past, followed
by negligible activity. They can be relatively easily selected from
the large SDSS spectral database by requiring no emission in H$\alpha$
or in [OII] and equivalent width of H$\delta > 5$\AA~ in absorption
(following \cite[Goto 2005]{goto05}). In addition we require
light-weighted mean stellar age between 0.5 and 1.5 Gyr as determined
from optical spectral indexes (little sensitive to TP-AGB modeling)
according to the bayesian method of \cite[Gallazzi
\etal~(2005)]{gallazzi+05}. As the most prominent NIR features of
TP-AGB stars are located at 1.41 and 1.77\micron~and, thus, fall in
the atmospheric gaps between J and H, and H and K bands for $z\sim0$
galaxies, we selected galaxies at $z\sim$0.2, so that those two
features move to the middle of the H and K atmospheric windows,
respectively. The resulting sample of 16 galaxies provides a good
coverage of the age-metallicity domain that is most strongly affected
by TP-AGB stars.

\begin{figure}[t]
\vspace*{-1.6 cm}
\begin{center}
 \includegraphics[width=0.75\textwidth]{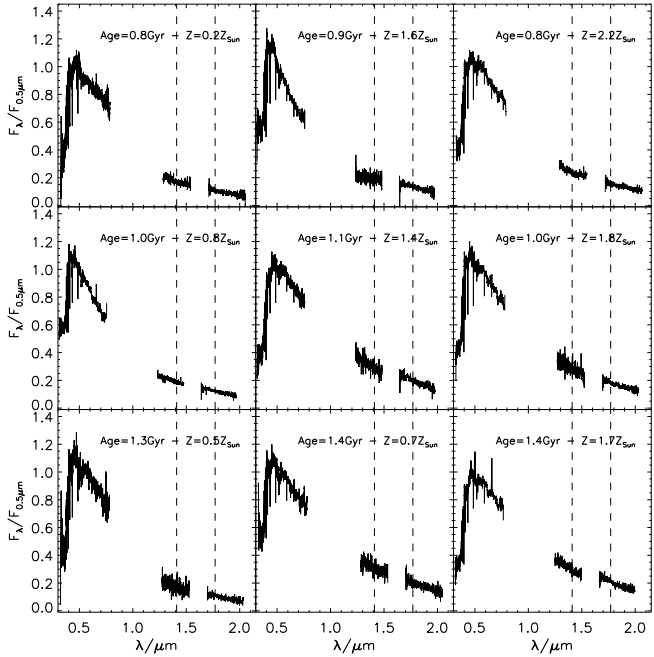} 
 \vspace*{-1.2 cm}
 \caption{Nine examples of combined optical-NIR spectra (ISAAC for the
   NIR, SDSS for the optical) of the post-starburst galaxies in our
   sample. The three rows show galaxies of increasing mean
   light-weighted age, from $\sim$0.8 Gyr (top), to $\sim$1 Gyr
   (middle) to $\sim$1.4 Gyr (bottom). In each row light-weighted mean
   metallicity increases from left to right, from sub-solar to
   super-solar values.  The spectra in each panel are cross-calibrated
   based on SDSS, UKIDSS and ISAAC photometry normalized to the flux
   at 0.5\micron (rest) and shown as a function of rest-frame
   wavelength. The vertical dashed lines mark 1.41 and 1.77\micron,
   where sharp spectral features are expected based on the Maraston
   (2005) models (cf. their Fig. 14, 15 and 18). However neither is
   observed.  Each panel is labeled with the light-weighted stellar
   age and metallicity derived using the bayesian method of Gallazzi
   et al. (2005).}
   \label{fig:obs9pan}
\end{center}
\end{figure}

NIR spectra in H and K bands have been obtained with ISAAC at the
ESO-VLT during P86 and assembled with the SDSS spectra to provide
flux-calibrated SEDs ranging from $\sim$3000~to $\sim$21000\AA~
with a rest-frame resolution of 3\AA~in the optical and 40\AA~ in the
NIR and typical S/N$>$30 per resolution element (see examples in
Fig. \ref{fig:obs9pan}). Spectra have been rescaled to match the
broad-band ``total'' fluxes derived from SDSS, UKIDSS and ISAAC
imaging in matched apertures corresponding to the SDSS Petrosian
aperture.

{\underline{\it Where are the TP-AGB NIR spectral features?}}  The H-
and K-band spectra of all 16 galaxies in the sample appear smooth and
featureless. In particular we do not detect either of the two sharp
features at 1.41\micron~and 1.77\micron~(C$_2$) rest-frame, in clear
contrast with the Maraston (2005) models predicting extremely strong
and abrupt flux drops of 10 to 30\% at those wavelengths. Our
observations are the first ones to prove this \textit{directly} thanks
to the NIR spectroscopy and the choice of a suitable redshift range.
Nine examples out of our 16 observed galaxies are reported in
Fig. \ref{fig:obs9pan} to demonstrate the lack of such NIR spectral
features. Spanning the full range of metallicities over the age range
where TP-AGB stars are expected to peak, there is no sign of sharp NIR
spectral features.

{\underline{\it ``TP or not TP? that is the question''}} Over the last
few years a general consensus has grown among SPS modelists about the
need for a boosted NIR flux due to TP-AGB stars with respect to
earlier generation models (e.g. BC03), although with strong
differences concerning its exact amount. \cite[Kriek
\etal~(2010)]{kriek+10} have recently challenged this, by showing that
the composite (medium-band NIR) SED of 62 post-starburst galaxies at
$0.7<z< 2.0$ is fully consistent with BC03 and inconsistent with
Maraston (2005). Although very careful analysis is needed to compare
our new observations with different models (in prep.), preliminary
analysis shows that the majority of our spectra is extremely well
reproduced by BC03 simple stellar populations (SSP) with no need for
further tweaking/fitting or inclusion of dust. This suggests that
predictions from early models might not require very substantial
correction in terms of NIR output.

\end{document}